\documentclass[showpacs,aps,graphicx]{revtex4}
\usepackage{graphicx}

\begin{document}
\title{The effective protection protocol of single photon state from photon loss and decoherence}

\author{Lan Zhou$^{1,2}$, Yu-Bo Sheng,$^{2,3}$\footnote{Email address:
shengyb@njupt.edu.cn}  }
\address{
 $^1$ College of Mathematics \& Physics, Nanjing University of Posts and Telecommunications, Nanjing,
210003, China\\
 $^2$Key Lab of Broadband Wireless Communication and Sensor Network
 Technology, Nanjing University of Posts and Telecommunications, Ministry of
 Education, Nanjing, 210003, China\\
$^3$Institute of Signal Processing  Transmission, Nanjing
University of Posts and Telecommunications, Nanjing, 210003,  China\\}

\begin{abstract}
We design an effect protocol for protecting the single-photon entanglement from photon loss and decoherence. The protocol only requires some auxiliary single photons and the linear optical elements. By operating the protocol, the photon loss can be effectively decreased and the less entangled  single photon state can be recovered to the maximally entangled state with some probability. Moreover, the polarization information encoded in the single photon state can be perfectly contained. The protocol can be realized under current experimental condition. As the single photon entanglement is quite important in quantum communication, this protocol may be useful in current and future quantum information processing.
\end{abstract}
 \maketitle

\section{Introduction}
Entanglement is a unique phenomenon of quantum mechanics, which was first described by Einstein, Podolsky, and Rosen \cite{entanglement1} and $Schr\ddot{o}dinger$ \cite{entanglement2}. Over the past few decades, together with the rapid experimental progress on quantum control, entanglement was widely used as a resource, enabling tasks like quantum teleportation \cite{teleportation}, quantum key distribution (QKD) \cite{qkd},  quantum secret sharing (QSS) \cite{qss}, quantum secure direct communication (QSDC) \cite{qsdc1,qsdc2}, quantum repeaters \cite{repeater1,repeater2}, and one-way quantum computation \cite{oneway,zhoucluster,shengcluster}. In the fields of quantum communication and computation, the single-photon entanglement (SPE) with the form of $\frac{1}{\sqrt{2}}(|1\rangle_{A}|0\rangle_{B}+|0\rangle_{A}|1\rangle_{B})$ is a simple but quite important entanglement form \cite{SPE1,SPE2,SPE3,SPE4}. It means that the single photon may be in location A or in location B with the same probability. $|0\rangle$ and $|1\rangle$ mean no photon and one photon, respectively. The SPE has been regarded as a valuable resource
for cryptography \cite{cryptography1,cryptography2}, state engineering \cite{engineering}, and tomography
of states and operations \cite{tomography1,tomography2}. Especially, the SPE has important applications in the quantum repeater protocols of long-distance quantum communication. For example, in the well known Duan-Lukin-Cirac-Zoller (DLCZ) protocol, the quantum repeater was realized based on the SPE and
cold atomic ensembles \cite{DLCZ}. In 2012, the group of Gottesman proposed a protocol for
building an interferometric telescope with the SPE \cite{telescope}. The protocol has the potential to eliminate the
baseline length limit, and allows the interferometers
with arbitrarily long baselines in principle.

Unfortunately, like other entanglement forms, SPE is sensitive to environmental noise during the transmission and storage process. The environmental noise may cause the photon loss or entanglement decoherence. The photon loss is one of the main obstacles for long-distance
quantum communication. It will make the pure state mix with the vacuum state with some probability. The photon loss will cause the degree of entanglement between two distant sites normally decreases exponentially
with the length of the connecting channel \cite{DLCZ}. On the other hand, the decoherence may make the maximally entangled state degrade to the mixed state or the pure less entangled state. These degraded states can not be used to setup the high-quality quantum entanglement channel, worse still, they will cause the communication insecure \cite{DLCZ}. Therefore, in practical applications, both the photon loss and decoherence obstacles need to be solved. The photon noiseless linear amplification (NLA) is the powerful way to overcome the photon loss, which was first proposed by Ralph and Lund \cite{NLA1}. Subsequently, various interesting NLA protocols have been proposed \cite{Gisin,Xiang,Pitkanen,Curty,heralded,heralded2,heralded3,Zhang,zhoul,wangtj,wangtj2,sheng2,ouyang,zhouamp,Evan,ou,feng}. For example, in 2010, Gisin \emph{et al.} investigated the device-independent QKD based on the NLA \cite{Gisin}. In the same year, Xiang \emph{et al.} also realized the heralded NLA and the
distillation of entanglement \cite{Xiang}. In 2012, Osorio \emph{et al.} showed their experimental
results of heralded quantum amplifier based on single-photon sources and
linear optics \cite{heralded}. The experiment of heralded noiseless amplification
of a photon polarization qubit was also reported soon later \cite{heralded2}. In 2013, the group of Evan proposed NLA, which can not only increase the fidelity of single photon, but also protect the encoded polarization information with some probability \cite{Evan}. In 2015, Sheng \emph{et al.} also proposed the recyclable amplification protocol for the SPE, which increases the fidelity of the single-photon state repeatedly. \cite{zhouamp}. Recently, Bruno
\emph{et al.} proposed a simple heralded qubit amplification protocol for the time-bin qubit based on linear optics. They showed this protocol has potential for fully integrated photonic solutions \cite{timebin}. On the other hand, the method to recover the pure less entangled state into the maximally entangled state is called the entanglement concentration. In 1996, Bennett \emph{et al.} first proposed the concept of entanglement concentration for the two-particle less-entangled state \cite{Bennett1}.
Since then, various interesting
ECPs have been put forward, successively \cite{swapping1,swapping2,zhao1,Yamamoto1,shengpra2,shengqic,shengpra3,shengwstateconcentration,bose,wangxb,wangc2,dengpra,yeliu,gubin,zhounoon,zhounoon2,zhouqip2,zhouoc}. In
2001, Zhao \emph{et al.} and Yamamoto \emph{et al.} proposed two similar
concentration protocols based on polarizing beam splitters
(PBSs), and realized them independently \cite{zhao1,Yamamoto1}. In 2008, Sheng \emph{et al.} developed
their protocols with the help of the cross-Kerr nonlinearity \cite{shengpra2}. Later, Sheng \emph{et al.} firstly proposed the ECP for the SPE \cite{shengqic} and a NLA for protecting the single-photon entanglement from both the photon loss and decoherence \cite{zhoul}.

Inspired by the research work from Ref. \cite{timebin} and our previous NLA protocols, in the paper, we design an effective NLA protocol for the SPE. This NLA protocol not only can protect the SPE from both the photon loss and decoherence, but also can keep the encoded polarization information of the SPE. The NLA protocol only requires some auxiliary single photons and the linear optical elements, such as the variable beam splitter (VBS), and 50:50 beam splitter (BS), which makes it can be realized under current experimental conditions. The protocol may be useful in current and future quantum information processing. This paper is organized as follows: In Sec. 2, we explain the basic principles of the amplification and concentration protocol. In Sec. 3, we make a discussion and summary.

\section{The amplification protocol for the SPE state }
\begin{figure}[!h]
\begin{center}
\includegraphics[width=12cm,angle=0]{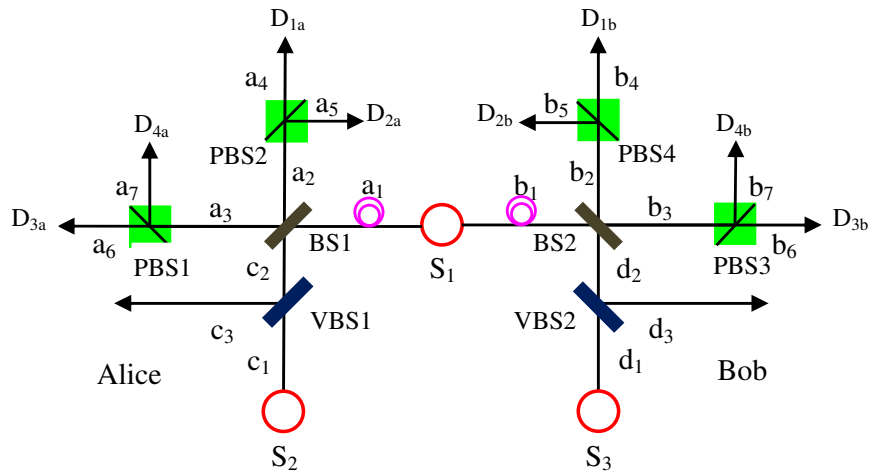}
\caption{The schematic principle of the amplification  protocol for the SPE. The SPE  is generated from $S1$. The auxiliary single photons are generated in the single photon source $S2$ and $S3$, respectively. VBS1 and VBS2 present the varable beam splitters with the transmission of $t_{1}$ and $t_{2}$, respectively. BSs mean 50:50 beam splitters. The PBSs mean the polarizing beam splitters, which can totally transmit the photon in $|H\rangle$ but totally reflect the photon in $|V\rangle$. $D_{1a}$, $D_{2a}$, $D_{3a}$, $D_{4a}$, $D_{1b}$, $D_{2b}$, $D_{3b}$, and $D_{4b}$ are single photon detectors.  }
\end{center}
\end{figure}In the protocol, we explain our protocol for protecting the SPE assisted with some auxiliary single photons in detail. The basic principle of the protocol is shown in Fig. 1. Suppose the single photon source $S_{1}$ generates a single-photon qubit with the form of
\begin{eqnarray}
|\psi\rangle=\alpha|H\rangle+\beta|V\rangle,
\end{eqnarray}
where $\alpha$ and $\beta$ are real, and $\alpha^{2}+\beta^{2}=1$. The single-photon qubit is sent to two parties, say Alice and Bob, which creates a maximally entangled SPE state as
\begin{eqnarray}
|\Phi\rangle_{AB}=\frac{1}{\sqrt{2}}(|\psi\rangle_{a1}|0\rangle_{b1}+|0\rangle_{a1}|\psi\rangle_{b1}).\label{single}
\end{eqnarray}
Due to the channel noise, the single photon entangled state can be degraded to a mixed state with the form of
\begin{eqnarray}
\rho_{in}=\eta|\Phi'\rangle_{AB}\langle\Phi'|+(1-\eta)|vac\rangle\langle vac|.
\end{eqnarray}
The fidelity $\eta$ shows that the single photon may be completely lost with the probability of $1-\eta$. Here, $|\Phi'\rangle_{AB}$ is a less entangled state as
\begin{eqnarray}
|\Phi'\rangle_{AB}=a|\psi\rangle_{a1}|0\rangle_{b1}+b|0\rangle_{a1}|\psi\rangle_{b1},
\end{eqnarray}
where $a$ and $b$ are two real entanglement coefficients, and $a^{2}+b^{2}=1$. Our aim is to increase the fidelity of the mixed state, and make the entanglement coefficients $a$ and $b$ be the same.

For realizing the task, each of the two parties requires to prepare two auxiliary single photons. As shown in Fig. 1, both the single photon sources $S_{2}$ and $S_{3}$ emit two single photons in the spatial modes of $c_{1}$ and $d_{1}$, respectively. One photon is in $|H\rangle$, and other one is in $|V\rangle$. Each of the parties makes the photons in his or her hand pass through a variable beam splitter (VBS), here named VBS1 and VBS2, respectively. The two VBSs have the transmission of $t_{1}$ and $t_{2}$, respectively. After the VBSs, the auxiliary photon state will evolve to
\begin{eqnarray}
|\varphi_{AB}\rangle&=&(\sqrt{t_{1}}|H\rangle_{c2}|0\rangle_{c3}+\sqrt{1-t_{1}}|0\rangle_{c2}|H\rangle_{c3})\otimes(\sqrt{t_{1}}|V\rangle_{c2}|0\rangle_{c3}+\sqrt{1-t_{1}}|0\rangle_{c2}|V\rangle_{c3})\nonumber\\
&\otimes&(\sqrt{t_{2}}|H\rangle_{d2}|0\rangle_{d3}+\sqrt{1-t_{2}}|0\rangle_{d2}|H\rangle_{d3})\otimes(\sqrt{t_{2}}|V\rangle_{d2}|0\rangle_{d3}+\sqrt{1-t_{2}}|0\rangle_{d2}|V\rangle_{d3})\nonumber\\
&=&[t_{1}|HV\rangle_{c2}+(1-t_{1})|HV\rangle_{c3}+\sqrt{t_{1}(1-t_{1})}(|H\rangle_{c2}|V\rangle_{c3}+|V\rangle_{c2}|H\rangle_{c3})]\nonumber\\
&\otimes&[t_{2}|HV\rangle_{d2}+(1-t_{2})|HV\rangle_{d3}+\sqrt{t_{2}(1-t_{2})}(|H\rangle_{d2}|V\rangle_{d3}+|V\rangle_{d2}H\rangle_{d3})].
\end{eqnarray}

 Considering the initial mixed state, the whole state $\rho_{in}\otimes|\varphi_{AB}\rangle$ can be described as follows. It is in the state of $|\Phi'\rangle_{AB}\otimes|\varphi_{AB}\rangle$ with the probability of $\eta$, or in the state of $|vac\rangle\otimes|\varphi_{AB}\rangle$ with the probability of $1-\eta$. We first discuss the case of $|\Phi'\rangle_{AB}\otimes|\varphi_{AB}\rangle$. The whole photon state can be written as
\begin{eqnarray}
&&|\Phi'\rangle_{AB}\otimes|\varphi_{AB}\rangle
=[a(\alpha|H\rangle_{a1}+\beta|V\rangle_{a1})|0\rangle_{b1}
+b|0\rangle_{a1}(\alpha|H\rangle_{b1}+\beta|V\rangle_{b1})]\nonumber\\
&\otimes&[t_{1}|HV\rangle_{c2}+(1-t_{1})|HV\rangle_{c3}+\sqrt{t_{1}(1-t_{1})}(|H\rangle_{c2}|V\rangle_{c3}+|V\rangle_{c2}|H\rangle_{c3})]\nonumber\\
&\otimes&[t_{2}|HV\rangle_{d2}+(1-t_{2})|HV\rangle_{d3}+\sqrt{t_{2}(1-t_{2})}(|H\rangle_{d2}|V\rangle_{d3}+|V\rangle_{d2}H\rangle_{d3})].\label{whole1}
\end{eqnarray}

Then, the two parties make the photons in the $a_{1}c_{2}$ and $b_{1}d_{2}$ modes pass through two $50:50$ beam splitters (BSs), here named BS1 and BS2, respectively. The BSs can make
\begin{eqnarray}
|1\rangle_{a1}&=&\frac{1}{\sqrt{2}}(|1\rangle_{a2}+|1\rangle_{a3}),\quad |1\rangle_{c2}=\frac{1}{\sqrt{2}}(|1\rangle_{a2}-|1\rangle_{a3}),\nonumber\\
|1\rangle_{b1}&=&\frac{1}{\sqrt{2}}(|1\rangle_{b2}+|1\rangle_{b3}),\quad |1\rangle_{d2}=\frac{1}{\sqrt{2}}(|1\rangle_{b2}-|1\rangle_{b3}).
\end{eqnarray}
After the BSs, the state in Eq. (\ref{whole1}) will evolve to
\begin{eqnarray}
|\Phi\rangle_{AB}\otimes|\varphi_{AB}\rangle&\rightarrow&\frac{1}{\sqrt{2}}\{(a\alpha|H\rangle_{a2}+a\alpha|H\rangle_{a3}+a\beta|V\rangle_{a2}+a\beta|V\rangle_{a3})\nonumber\\
&\otimes&[\frac{t_{1}}{2}(|HV\rangle_{a2}-|H\rangle_{a2}|V\rangle_{a3}-|V\rangle_{a2}|H\rangle_{a3}+|HV\rangle_{a3})+(1-t_{1})|HV\rangle_{c3}\nonumber\\
&+&\frac{\sqrt{t_{1}(1-t_{1})}}{\sqrt{2}}(|H\rangle_{a2}|V\rangle_{c3}-|H\rangle_{a3}|V\rangle_{c3}-|V\rangle_{a2}|H\rangle_{c3}-|V\rangle_{a3}|H\rangle_{c3})]\nonumber\\
&\otimes&[\frac{t_{2}}{2}(|HV\rangle_{b2}-|H\rangle_{b2}|V\rangle_{b3}-|V\rangle_{b2}|H\rangle_{b3}+|HV\rangle_{b3})
+(1-t_{2})|HV\rangle_{d3}\nonumber\\
&+&\frac{\sqrt{t_{2}(1-t_{2})}}{\sqrt{2}}(|H\rangle_{b2}|V\rangle_{d3}-|H\rangle_{b3}|V\rangle_{d3}-|V\rangle_{b2}|H\rangle_{d3}-|V\rangle_{b3}|H\rangle_{d3})]\nonumber\\
&+&(b\alpha|H\rangle_{b2}+b\alpha|H\rangle_{b3}+b\beta|V\rangle_{b2}+b\beta|V\rangle_{b3})\nonumber\\
&\otimes&[\frac{t_{1}}{2}(|HV\rangle_{a2}-|H\rangle_{a2}|V\rangle_{a3}-|V\rangle_{a2}|H\rangle_{a3}+|HV\rangle_{a3})
+(1-t_{1})|HV\rangle_{c3}\nonumber\\
&+&\frac{\sqrt{t_{1}(1-t_{1})}}{\sqrt{2}}(|H\rangle_{a2}|V\rangle_{c3}-|H\rangle_{a3}|V\rangle_{c3}-|V\rangle_{a2}|H\rangle_{c3}-|V\rangle_{a3}|H\rangle_{c3})]\nonumber\\
&\otimes&[\frac{t_{2}}{2}(|HV\rangle_{b2}-|H\rangle_{b2}|V\rangle_{b3}-|V\rangle_{b2}|H\rangle_{b3}+|HV\rangle_{b3})
+(1-t_{2})|HV\rangle_{d3}\nonumber\\
&+&\frac{\sqrt{t_{2}(1-t_{2})}}{\sqrt{2}}(|H\rangle_{b2}|V\rangle_{d3}-|H\rangle_{b3}|V\rangle_{d3}-|V\rangle_{b2}|H\rangle_{d3}-|V\rangle_{b3}|H\rangle_{d3})]\}.\label{whole2}
\end{eqnarray}

Then, the parties make the Bell-state measurement for the output photons of the BSs. They make the photons in the $a_{2}a_{3}$ and $b_{2}b_{3}$ modes pass through four polarizing beam splitters (PBSs), here named $PBS1$, $PBS2$ and $PBS3$, $PBS4$, respectively. The PBS can totally transmit the photon in $|H\rangle$ but totally reflect the photon in $|V\rangle$. In this way, after the PBSs, the whole state in Eq. (\ref{whole2}) will evolve to
\begin{eqnarray}
&&\frac{1}{\sqrt{2}}\{(a\alpha|H\rangle_{a4}+a\alpha|H\rangle_{a6}+a\beta|V\rangle_{a5}+a\beta|V\rangle_{a7})\nonumber\\
&\otimes&[\frac{t_{1}}{2}(|H\rangle_{a4}|V\rangle_{a5}-|H\rangle_{a4}|V\rangle_{a7}-|V\rangle_{a5}|H\rangle_{a6}+|H\rangle_{a6}|V\rangle_{a7})+(1-t_{1})|HV\rangle_{c3}\nonumber\\
&+&\frac{\sqrt{t_{1}(1-t_{1})}}{\sqrt{2}}(|H\rangle_{a4}|V\rangle_{c3}-|H\rangle_{a6}|V\rangle_{c3}-|V\rangle_{a5}|H\rangle_{c3}-|V\rangle_{a7}|H\rangle_{c3})]\nonumber\\
&\otimes&[\frac{t_{2}}{2}(|H\rangle_{b4}|V\rangle_{b5}-|H\rangle_{b4}|V\rangle_{b7}-|V\rangle_{b5}|H\rangle_{b6}+|H\rangle_{b6}|V\rangle_{b7})
+(1-t_{2})|HV\rangle_{d3}\nonumber\\
&+&\frac{\sqrt{t_{2}(1-t_{2})}}{\sqrt{2}}(|H\rangle_{b4}|V\rangle_{d3}-|H\rangle_{b6}|V\rangle_{d3}-|V\rangle_{b5}|H\rangle_{d3}-|V\rangle_{b7}|H\rangle_{d3})]\nonumber\\
&+&(b\alpha|H\rangle_{b4}+b\alpha|H\rangle_{b6}+b\beta|V\rangle_{b5}+b\beta|V\rangle_{b7})\nonumber\\
&\otimes&[\frac{t_{1}}{2}(|H\rangle_{a4}|V\rangle_{a5}-|H\rangle_{a4}|V\rangle_{a7}-|V\rangle_{a5}|H\rangle_{a6}+|H\rangle_{a6}|V\rangle_{a7})+(1-t_{1})|HV\rangle_{c3}\nonumber\\
&+&\frac{\sqrt{t_{1}(1-t_{1})}}{\sqrt{2}}(|H\rangle_{a4}|V\rangle_{c3}-|H\rangle_{a6}|V\rangle_{c3}-|V\rangle_{a5}|H\rangle_{c3}-|V\rangle_{a7}|H\rangle_{c3})]\nonumber\\
&\otimes&[\frac{t_{2}}{2}(|H\rangle_{b4}|V\rangle_{b5}-|H\rangle_{b4}|V\rangle_{b7}-|V\rangle_{b5}|H\rangle_{b6}+|H\rangle_{b6}|V\rangle_{b7})
+(1-t_{2})|HV\rangle_{d3}\nonumber\\
&+&\frac{\sqrt{t_{2}(1-t_{2})}}{\sqrt{2}}(|H\rangle_{b4}|V\rangle_{d3}-|H\rangle_{b6}|V\rangle_{d3}-|V\rangle_{b5}|H\rangle_{d3}-|V\rangle_{b7}|H\rangle_{d3})]\}.\label{whole3}
\end{eqnarray}
The photons in the eight output modes of the four PBSs are detected by the single-photon detectors, respectively. The eight single-photon detectors are called $D_{1a}$, $D_{2a}$, $D_{3a}$, $D_{4a}$, $D_{1b}$, $D_{2b}$, $D_{3b}$, and $D_{4b}$, respectively. According to the measurement results, the parties can ensure whether the amplification protocol is successful. Under the case that the detectors $D_{1a}D_{2a}$, $D_{1a}D_{4a}$, $D_{2a}D_{3a}$, or $D_{3a}D_{4a}$ in Alice's location each registers one photon, simultaneously, the detectors $D_{1b}D_{2b}$, $D_{1b}D_{4b}$, $D_{2b}D_{3b}$, or $D_{3b}D_{4b}$ in Bob's location each registers one photon, the amplification protocol will be successful. Otherwise, the protocol becomes a failure. In this way, there are sixteen successful measurement results in total, say, $D_{1a}D_{2a}D_{1b}D_{2b}$, $D_{1a}D_{2a}D_{1b}D_{4b}$, $D_{1a}D_{2a}D_{1b}D_{4b}$, $D_{1a}D_{2a}D_{1b}D_{4b}$, $D_{1a}D_{4a}D_{1b}D_{2b}$, $D_{1a}D_{4a}D_{1b}D_{4b}$, $D_{1a}D_{4a}D_{2b}D_{3b}$, $D_{1a}D_{4a}D_{3b}D_{4b}$, $D_{2a}D_{3a}D_{1b}D_{2b}$, $D_{2a}D_{3a}D_{1b}D_{4b}$, $D_{2a}D_{3a}D_{2b}D_{3b}$, $D_{2a}D_{3a}D_{3b}D_{4b}$, $D_{3a}D_{4a}D_{1b}D_{2b}$, $D_{3a}D_{4a}D_{1b}D_{4b}$, $D_{3a}D_{4a}D_{2b}D_{3b}$, $D_{3a}D_{4a}D_{3b}D_{4b}$.

Under all the sixteen successful measurement results, the parties can finally obtain the same output photon state. We take the measurement result of $D_{1a}D_{2a}D_{1b}D_{2b}$ and $D_{1a}D_{2a}D_{1b}D_{4b}$ for example. From Eq. (\ref{whole3}), it can be found that the four items $\frac{-a\alpha t_{2}\sqrt{t_{1}(1-t_{1})}}{4}|H\rangle_{a4}|V\rangle_{a5}|H\rangle_{b4}|V\rangle_{b5}|H\rangle_{c3}$, $\frac{a\beta t_{2}\sqrt{t_{1}(1-t_{1})}}{4}|V\rangle_{a5}|H\rangle_{a4}|H\rangle_{b4}|V\rangle_{b5}|V\rangle_{c3}$, $\frac{-b\alpha t_{1}\sqrt{t_{2}(1-t_{2})}}{4}|H\rangle_{b4}|V\rangle_{b5}|H\rangle_{a4}|V\rangle_{a5}|H\rangle_{d3}$, and $\frac{b\beta t_{1}\sqrt{t_{2}(1-t_{2})}}{4}|V\rangle_{b5}|H\rangle_{b4}|H\rangle_{a4}|V\rangle_{a5}|V\rangle_{d3}$ will lead to the measurement result of $D_{1a}D_{2a}D_{1b}D_{2b}$. Therefore, under this measurement result, the whole photon state in Eq. (\ref{whole3}) will collapse to
\begin{eqnarray}
|\Phi'_{1}\rangle_{AB}=at_{2}\sqrt{t_{1}(1-t_{1})}(-\alpha|H\rangle_{c3}+\beta|V\rangle_{c3})|0\rangle_{d3}
+bt_{1}\sqrt{t_{2}(1-t_{2})}|0\rangle_{c3}(-\alpha|H\rangle_{d3}+\beta|V\rangle_{d3}),\label{result1}
\end{eqnarray}
with the probability of $\frac{a^{2}t_{1}t_{2}^{2}(1-t_{1})+b^{2}t_{1}^{2}t_{2}(1-t_{2})}{16}$.
Similarly, from Eq. (\ref{whole3}), all the four items $\frac{a\alpha t_{2}\sqrt{t_{1}(1-t_{1})}}{4}|H\rangle_{a4}|V\rangle_{a5}|H\rangle_{b4}|V\rangle_{b7}|H\rangle_{c3}$, $\frac{-a\beta t_{2}\sqrt{t_{1}(1-t_{1})}}{4}|V\rangle_{a5}|H\rangle_{a4}|H\rangle_{b4}|V\rangle_{b7}|V\rangle_{c3}$, $\frac{-b\alpha t_{1}\sqrt{t_{2}(1-t_{2})}}{4}|H\rangle_{b4}|V\rangle_{b7}|H\rangle_{a4}|V\rangle_{a5}|H\rangle_{d3}$, and $\frac{b\beta t_{1}\sqrt{t_{2}(1-t_{2})}}{4}|V\rangle_{b7}|H\rangle_{b4}|H\rangle_{a4}|V\rangle_{a5}|V\rangle_{d3}$ will lead to the measurement result of $D_{1a}D_{2a}D_{1b}D_{4b}$. In this way, under this measurement result, the whole photon state will collapse to
\begin{eqnarray}
|\Phi"_{1}\rangle_{AB}=at_{2}\sqrt{t_{1}(1-t_{1})}(\alpha|H\rangle_{c3}-\beta|V\rangle_{c3})|0\rangle_{d3}
-bt_{1}\sqrt{t_{2}(1-t_{2})}|0\rangle_{c3}(\alpha|H\rangle_{d3}-\beta|V\rangle_{d3}),\label{result2}
\end{eqnarray}
with the same probability. Obviously, both Eq. (\ref{result1}) and Eq. (\ref{result2}) can be transformed to
\begin{eqnarray}
|\Phi_{1}\rangle_{AB}=at_{2}\sqrt{t_{1}(1-t_{1})}(\alpha|H\rangle_{c3}+\beta|V\rangle_{c3})|0\rangle_{d3}
+bt_{1}\sqrt{t_{2}(1-t_{2})}|0\rangle_{c3}(\alpha|H\rangle_{d3}+\beta|V\rangle_{d3}), \label{result3}
\end{eqnarray}
by performing a phase-flip operation on the photon in the $c_{3}$ or $d_{3}$ mode. If the parties obtain one of other fourteen successful measurement results, they can also finally obtain $|\Phi_{1}\rangle_{AB}$ in Eq. (\ref{result3}). From Eq. (\ref{result3}), it can be found that the $|\Phi_{1}\rangle_{AB}$ can be transformed to the maximally entangled single-photon entangled state as $|\Phi\rangle_{AB}$ in Eq. (\ref{single}) when the transmission of the VBSs satisfies
\begin{eqnarray}
\frac{a^{2}}{b^{2}}=\frac{t_{1}(1-t_{2})}{t_{2}(1-t_{1})}.\label{require1}
\end{eqnarray}
In this way, as $a^{2}+b^{2}=1$, we can obtain $t_{2}$ as a function of $t_{1}$ and $a$, which can be written as
\begin{eqnarray}
t_{2}=\frac{t_{1}(1-a^{2})}{a^{2}-2a^{2}t_{1}+t_{1}}.\label{require1}
\end{eqnarray}

Therefore, the success probability of our protocol corresponding to the initial input state of $|\Phi'\rangle_{AB}$ as
\begin{eqnarray}
 P_{1}=16\times\frac{a^{2}t_{1}t_{2}^{2}(1-t_{1})+b^{2}t_{1}^{2}t_{2}(1-t_{2})}{16}=\frac{2a^{2}(1-a^{2})^{2}t_{1}^{3}(1-t_{1})}{(a^{2}-2a^{2}t_{1}+t_{1})^{2}}.
\end{eqnarray}

On the other hand, if the initial input state is the vacuum state, after the auxiliary photons pass through BS1 and BS2, respectively, the parties can obtain
\begin{eqnarray}
|vac\rangle\otimes|\varphi_{AB}\rangle&\rightarrow&[\frac{t_{1}}{2}(|HV\rangle_{a2}-|H\rangle_{a2}|V\rangle_{a3}-|V\rangle_{a2}|H\rangle_{a3}+|HV\rangle_{a3})+(1-t_{1})|HV\rangle_{c3}\nonumber\\
&+&\frac{\sqrt{t_{1}(1-t_{1})}}{\sqrt{2}}(|H\rangle_{a2}|V\rangle_{c3}-|H\rangle_{a3}|V\rangle_{c3}-|V\rangle_{a2}|H\rangle_{c3}-|V\rangle_{a3}|H\rangle_{c3})]\nonumber\\
&\otimes&[\frac{t_{2}}{2}(|HV\rangle_{b2}-|H\rangle_{b2}|V\rangle_{b3}-|V\rangle_{b2}|H\rangle_{b3}+|HV\rangle_{b3})
+(1-t_{2})|HV\rangle_{d3}\nonumber\\
&+&\frac{\sqrt{t_{2}(1-t_{2})}}{\sqrt{2}}(|H\rangle_{b2}|V\rangle_{d3}-|H\rangle_{b3}|V\rangle_{d3}-|V\rangle_{b2}|H\rangle_{d3}-|V\rangle_{b3}|H\rangle_{d3})].
\label{whole4}
\end{eqnarray}

Next, the parties also make the photons in the $a_{2}a_{3}b_{2}b_{3}$ modes pass through four PBSs, respectively. After the PBSs, the whole state in Eq. (\ref{whole4}) will evolve to
\begin{eqnarray}
|vac\rangle\otimes|\varphi_{AB}\rangle&\rightarrow&[\frac{t_{1}}{2}(|H\rangle_{a4}|V\rangle_{a5}-|H\rangle_{a4}|V\rangle_{a7}-|V\rangle_{a5}|H\rangle_{a6}+|H\rangle_{a6}|V\rangle_{a7})+(1-t_{1})|HV\rangle_{c3}\nonumber\\
&+&\frac{\sqrt{t_{1}(1-t_{1})}}{\sqrt{2}}(|H\rangle_{a4}|V\rangle_{c3}-|H\rangle_{a6}|V\rangle_{c3}-|V\rangle_{a5}|H\rangle_{c3}-|V\rangle_{a7}|H\rangle_{c3})]\nonumber\\
&\otimes&[\frac{t_{2}}{2}(|H\rangle_{b4}|V\rangle_{b5}-|H\rangle_{b4}|V\rangle_{b7}-|V\rangle_{b5}|H\rangle_{b6}+|H\rangle_{b6}|V\rangle_{b7})
+(1-t_{2})|HV\rangle_{d3}\nonumber\\
&+&\frac{\sqrt{t_{2}(1-t_{2})}}{\sqrt{2}}(|H\rangle_{b4}|V\rangle_{d3}-|H\rangle_{b6}|V\rangle_{d3}-|V\rangle_{b5}|H\rangle_{d3}-|V\rangle_{b7}|H\rangle_{d3})].
\label{whole5}
\end{eqnarray}
After the PBSs, the photons in the $a_{4}$, $a_{5}$, $a_{6}$, $a_{7}$ and $b_{4}$, $b_{5}$, $b_{6}$, $b_{7}$ modes are measured by the $D_{1a}$, $D_{2a}$, $D_{3a}$, $D_{4a}$, and $D_{1b}$, $D_{2b}$, $D_{3b}$, $D_{4b}$, respectively. It is easy to found that if the parties obtain one of the above sixteen successful measurement results, the whole state in Eq. (\ref{whole5}) will finally collapse to the vacuum state. For example, in Eq. (\ref{whole5}), the item $\frac{t_{1}t_{2}}{4}|H\rangle_{a4}|V\rangle_{a5}|H\rangle_{b4}|V\rangle_{b5}$ will make the detectors $D_{1a}D_{2a}D_{1b}D_{2b}$ each detects one photon. In this way, under the measurement result of $D_{1a}D_{2a}D_{1b}D_{2b}$, Eq. (\ref{whole5}) will finally collapse to the vacuum state. Therefore, under the case that the initial input state is the vacuum state, the total success probability of our protocol  is $P_{2}=16\times\frac{t_{1}^{2}t_{2}^{2}}{16}=t_{1}^{2}t_{2}^{2}=\frac{t_{1}^{4}(1-a^{2})^{2}}{(a^{2}-2a^{2}t_{1}+t_{1})^{2}}$.

Combined the two possible cases of the input state, the total success probability $P_{t}$ of our amplification protocol can be written as
\begin{eqnarray}
P_{t}&=&\eta P_{1}+(1-\eta)P_{2}=\eta \frac{2a^{2}(1-a^{2})^{2}t_{1}^{3}(1-t_{1})}{(a^{2}-2a^{2}t_{1}+t_{1})^{2}}+ (1-\eta)\frac{t_{1}^{4}(1-a^{2})^{2}}{(a^{2}-2a^{2}t_{1}+t_{1})^{2}}\nonumber\\
&=&\frac{t_{1}^{3}(1-a^{2})^{2}[2\eta a^{2}(1-t_{1})+(1-\eta)t_{1}]}{(a^{2}-2a^{2}t_{1}+t_{1})^{2}}.
\end{eqnarray}
When the protocol is successful, the parties can obtain a mixed state as
\begin{eqnarray}
\rho_{out}=\eta'|\Phi_{1}\rangle_{AB}\langle\Phi_{1}|+(1-\eta')|vac\rangle\langle vac|
\end{eqnarray}
with the fidelity
\begin{eqnarray}
\eta'=\frac{\eta P_{1}}{P_{t}}=\frac{2\eta a^{2}(1-t_{1})}{2\eta a^{2}(1-t_{1})+(1-\eta)t_{1}}.\label{fidelity}
\end{eqnarray}

It can be found that the output mixed state $\rho_{out}$ has the same form of $\rho_{in}$.
The fidelity $\eta'$ of the new mixed state depends on the initial fidelity $\eta$, the initial entanglement coefficient $a$, and the transmission $t_{1}$ of VBS1. We design the amplification factor $g$ as
\begin{eqnarray}
g\equiv\frac{\eta'}{\eta}=\frac{2a^{2}(1-t_{1})}{2\eta a^{2}(1-t_{1})+(1-\eta)t_{1}}. \label{g}
\end{eqnarray}
For realizing the amplification, we require $\eta'>\eta$, that is, $g>1$. It can be calculated that $g>1$ under the case of $t_{1}<\frac{2a^{2}}{1+2a^{2}}$. In this way, by providing suitable VBS1 and VBS2 with $t_{1}<\frac{2a^{2}}{1+2a^{2}}$, and $t_{2}=\frac{t_{1}(1-a^{2})}{a^{2}-2a^{2}t_{1}+t_{1}}$, we can realize the concentration and amplification for the single-photon entangled state, simultaneously.

\section{Discussion and conclusion}

\begin{figure}[!h]
\begin{center}
\includegraphics[width=8cm,angle=0]{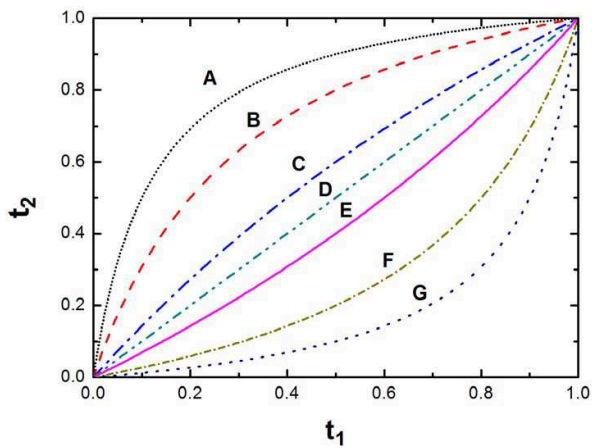}
\caption{The transmission $t_{2}$ as a function of $t_{1}$ under different initial value of $a$. Curve A, $a^{2}=0.1$; curve B, $a^{2}=0.2$; curve C, $a^{2}=0.4$; curve D, $a^{2}=0.5$; curve E, $a^{2}=0.6$; curve F, $a^{2}=0.8$; curve G, $a^{2}=0.9$.  }
\end{center}
\end{figure}

In the paper, we propose a simple and effective protocol for distilling the SPE from photon loss and decoherence with the help of some auxiliary single photons. In the protocol, each of the two parties requires to prepare two single photons with the polarization of $|H\rangle$ and $|V\rangle$. Then, Alice makes the photons in her location pass through the VBS1 with the transmission of $t_{1}$, and Bob makes the photons in his location pass through the VBS2 with the transmission of $t_{2}$, simultaneously. Subsequently, each party makes the photons enter the BS, and make the BSM for the photons in the output modes. According to the BSM result, the parties can distill a new mixed state with the similar form of the initial mixed state. By adjusting the coefficients $t_{1}$ and $t_{2}$ to meet Eq. (\ref{require1}), we can recover the less-entangled single-photon state into the maximally entangled single-photon state while keeping the initial polarization feature of the single photon. Moreover, we can obtain the amplification factor $g>1$ under the condition of $t_{1}<\frac{2a^{2}}{1+2a^{2}}$. In this way, we can realize the concentration and amplification for the SPE simultaneously by providing suitable VBSs.

For realizing the concentration, the values of $t_{2}$ as a function of $t_{1}$ under different values of $a^{2}$ are shown in Fig. 2. In Fig. 2, curve A, B, C, D, E, F, and G are corresponding to $a^{2}=0.1, 0.2, 0.4, 0.5, 0.6, 0.8$, and  $0.9$, respectively. From Eq. (\ref{require1}) and Fig. 2, it can be found if $a^{2}=\frac{1}{2}$ (curve D), that is, the initial single-photon state is the maximally entangled state, we will obtain $t_{2}=t_{1}$. Under other values of $a^{2}$, $t_{2}\neq t_{1}$. $t_{2}$ increases with the growth of $t_{1}$. When $t_{1}\rightarrow 0$, we can also obtain $t_{2}\rightarrow 0$.

\begin{figure}[!h]
\begin{center}
\includegraphics[width=8cm,angle=0]{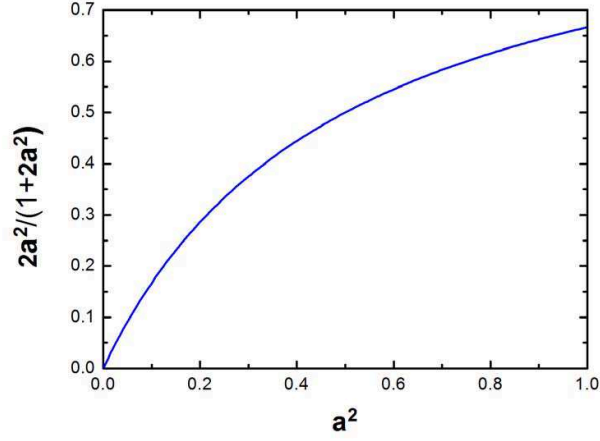}
\caption{The value of $\frac{2a^{2}}{1+2a^{2}}$ as a function of $a^{2}$. For realizing the amplification, we require $t_{1}<\frac{2a^{2}}{1+2a^{2}}$. In this way, the suitable value of $t_{1}$ is below the curve. }
\end{center}
\end{figure}

\begin{figure}[!h]
\begin{center}
\includegraphics[width=12cm,angle=0]{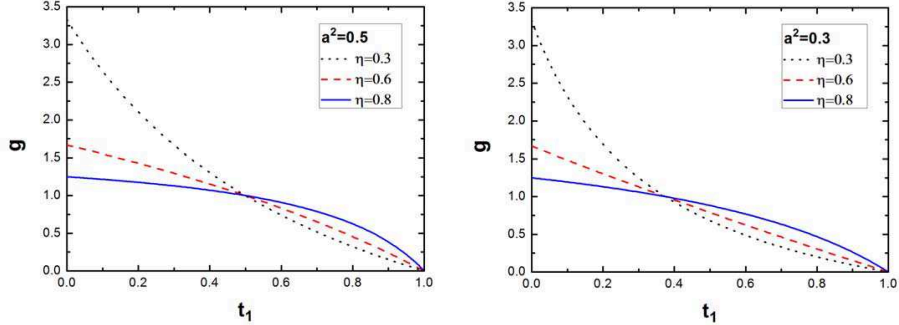}
\caption{The amplification factor $g$ as a function of $t_{1}$ under different entanglement coefficients and initial fidelity conditions. (a) $a^{2}=0.5$, $\eta=$0.3, 0.6, and 0.8, respectively. (b) $a^{2}=0.3$, $\eta=$0.3, 0.6, and 0.8, respectively. }
\end{center}
\end{figure}

\begin{figure}[!h]
\begin{center}
\includegraphics[width=12cm,angle=0]{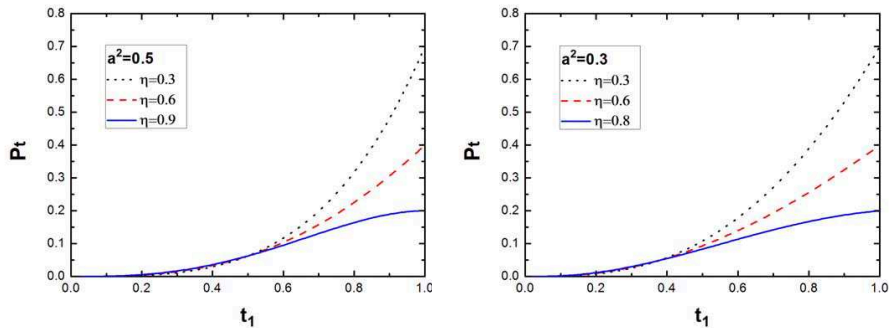}
\caption{The total success probability $P_{t}$ as a function of $t_{1}$ under different entanglement coefficients and initial fidelity conditions. (a) $a^{2}=0.5$, $\eta=$0.3, 0.6, and 0.8, respectively. (b) $a^{2}=0.3$, $\eta=$0.3, 0.6, and 0.8, respectively.}
\end{center}
\end{figure}

Fig. 3 and Fig. 4 show the value of $\frac{2a^{2}}{1+2a^{2}}$ as a function of $a^{2}$, and the amplification factor $g$ altered with $t_{1}$ under different $\eta$ and $a^{2}$, respectively. It can be found that $a^{2}$ controls the required scale of $t_{1}$. For realizing the amplification ($g>1$), we require $t_{1}<\frac{2a^{2}}{1+2a^{2}}$. Therefore, the required value of $t_{1}$ must be below the curve of Fig. 3. The higher value of $a^{2}$ leads to the broader range of $t_{1}$. For example, when $a^{2}=\frac{1}{2}$, we can obtain $t_{1}<\frac{1}{2}$, while when $a^{2}\rightarrow 1$, we can obtain $t_{1}<\frac{2}{3}$. From Fig. 4, when $t_{1}=\frac{2a^{2}}{1+2a^{2}}$, we can obtain $g=1$, which has nothing to do with the initial fidelity of $\eta$. Actually, under the case of $t=\frac{2a^{2}}{1+2a^{2}}$, our protocol is analogous to an entanglement concentration and swapping protocol. For obtaining high $g$, the parties need to prepare VBSs with relatively low transmission. Combined with Eq. (\ref{g}), we can obtain the limitation of $g$ as
\begin{eqnarray}
\lim_{t_{1}\rightarrow 0}g=\lim_{t_{1}\rightarrow 0}\frac{2a^{2}(1-t_{1})}{2\eta a^{2}(1-t_{1})+(1-\eta)t_{1}}=\frac{1}{\eta},
\end{eqnarray}
which has nothing to do with the value of $a^{2}$.

Fig. 5 shows the total success probability $P_{t}$ of our protocol as a function of $t_{1}$ under different $\eta$ and $a^{2}$. It can be found under the scale of $t_{1}<\frac{2a^{2}}{1+2a^{2}}$, the factors $\eta$ and $a^{2}$ have extremely slight effect on the $P_{t}$. $P_{t}$ mainly depends on the value of $t_{1}$. The lower value of $t_{1}$ leads to lower $P_{t}$. In this way, in practical applications, the parties need to consider the transmission requirements for both $g$ and $P_{t}$, and select the VBSs with suitable transmission.

Finally, we briefly discuss the experimental realization of our protocol. The VBS is the key element of the protocol. The parties complete the concentration and amplification tasks relying on adjusting the transmission of VBS1 and VBS2, respectively. They require to select the VBS1 with $t_{1}<\frac{2a^{2}}{1+2a^{2}}$, and the VBS2 with $t_{2}=\frac{t_{1}(1-a^{2})}{a^{2}-2a^{2}t_{1}+t_{1}}$. The VBS is a common linear optical
element in current technology, which has been widely used in current quantum communication field. For example, in 2012, the group of Osorio reported their experimental results
about the heralded photon amplification for quantum communication
with the help of the VBS \cite{heralded}. In their amplification experiment, they successfully adjusted the splitting
ratio of VBS from $50:50$ to $90:10$ to increase the visibility from
$46.7\pm 3.1\%$ to $96.3\pm 3.8\%$. Based on their experimental result, our requirement for the VBS can also be realized. On the other hand, we also require the sophisticated single photon detectors to exactly distinguish
the single photon in each output modes. Due to the quantum decoherence effect of the photon detector, the single photon detection has been a challenge under current experimental
conditions \cite{photonefficiency}. In 2008, Lita \emph{et al.} reported their experimental
result about the near-infrared single-photon detection. In their reports, the photon detection efficiency $\eta_{p}$ at 1556 $nm$ can reach $95\% \pm 2\%$ \cite{photonefficiency1}.

In conclusion, we design an effect protocol for protecting the single-photon entanglement from photon loss and decoherence. In the protocl, we only require some auxiliary single photons. By making photons pass through the VBSs and BSs, and adjusting the transmission of VBSs, we can realize the entanglement concentration and amplification simultaneously. Moreover, the polarization feature of the SPE can be perfectly contained. As the protocol only requires the linear optical elements, it can be realized under current experimental condition. In the protocol, the limitation of the amplification factor $g$ only depends on the fidelity $\eta$ of the initial mixed state, and we can obtain high value of $g$ under relatively low transmission. As the single photon entanglement is quite important in current long-distance quantum communication,
this protocol may be useful in current and future quantum information processing.


\begin{thebibliography}{99}
\bibitem{entanglement1}  Einstein, A.,  Podolsky, B.,  Rosen N.: Can quantum-mechanical description of physical reality be considered complete? Phys. Rev. \textbf{47}, 777 (1935)
\bibitem{entanglement2} Schr$\ddot{o}$dinger, E.: Die gegenwartige situation in der quantenmechanik. Die Naturwissenschaften \textbf{23}, 807-812 (1935)

\bibitem{teleportation}  Bennett, C.H.,  Brassard, G., Crepeau, C., Jozsa, R., Peres, A.,
 Wootters, W.K.: Teleporting an unknown quantum state via dual classical and Einstein-Podolsky-Rosen channels. Phys. Rev. Lett. \textbf{70}, 1895 (1993)

\bibitem{qkd} Ekert, A.K.: Quantum cryptography based on Bells theorem, Phys. Rev. Lett. \textbf{67}, 661-663 (1991)

\bibitem{qss} Hillery, M.,  Bu\v{z}ek, V., Berthiaume, A.: Quantum secret sharing. Phys. Rev. A \textbf{59}, 1829 (1999)

\bibitem{qsdc1}  Long, G.L., Liu, X.S.: Theoretically efficient high-capacity quantum-keydistribution scheme. Phys. Rev. A \textbf{65}, 032302 (2002)

\bibitem{qsdc2}  Deng, F.G.,  Long, G.L., Liu, X.S.: Two-step quantum direct communication protocol using the Einstein-Podolsky-Rosen pair
block. Phys. Rev. A \textbf{68}, 042317 (2003)

\bibitem{repeater1}  Briegel, H.J., D\"{u}r, W.,  Cirac, J.I., Zoller, P.: Quantum repeaters: the role of imperfect local operations in quantum communication. Phys. Rev. Lett. \textbf{81}, 5932 (1998)

\bibitem{repeater2} Sangouard, N., Simon, C., de Riedmatten, H., Gision, N.: Quantum repeaters based on atomic ensembles and linear optics. Rev. Mod. Phys. \textbf{83}, 33-80 (2011)

\bibitem{oneway}  Nielsen, M.A.: Optical quantum computation using cluster states. Phys. Rev. Lett. \textbf{93}, 040503 (2004)

\bibitem{zhoucluster} Zhou, L., Sheng, Y.B.: Arbitrary atomic cluster state concentration for one-way quantum computation. J. Opt. Soc. Am. B \textbf{31}, 503-511 (2014)

\bibitem{shengcluster}  Zhao, S.Y., Liu, J., Zhou, L., Sheng, Y.B.: Two-step entanglement concentration for arbitrary electronic cluster state. Quan. Inform. Process. \textbf{12}, 3633-3647 (2013)

\bibitem{SPE1} Lee, H.W., Kim, J.: Quantum teleportation and Bells inequality using single-particle entanglement. Phys. Rev. A \textbf{63}, 012305 (2001)
\bibitem{SPE2} van Enk, S.J.: Single-particle entanglement. Phys. Rev. A \textbf{72}, 064306 (2005)
\bibitem{SPE3} Hessmo, B., Usachev, P., Heydari, H., Bj$\ddot{o}$rk, G.: Experimental demonstration of single photon nonlocality. Phys. Rev.
Lett. \textbf{92}, 180401 (2004)

\bibitem{SPE4} Lombardi, E., Sciarrino, F., Popescu, S., De Martini, F.: Teleportation of a vacuum-one-photon qubit.
Phys. Rev. Lett. \textbf{88}, 070402 (2002)

\bibitem{cryptography1} Silberhorn, C., Ralph, T. C., L$\ddot{u}$tkenhaus, N., Leuchs, G.: Continuous variable quantum cryptography: beating the 3 dB loss limit. Phys. Rev. Lett. \textbf{89}, 167901 (2002)
\bibitem{cryptography2} Silberhorn, C., Korolkova, N., Leuchs, G.: Quantum key distribution with bright entangled beams. Phys. Rev.
Lett. \textbf{88}, 167902 (2002)

\bibitem{engineering} Paris, M.G.A., Cola, M., Bonifacio, R.: Quantum-state engineering assisted by entanglement. Phys. Rev. A
\textbf{67}, 042104 (2003)
\bibitem{tomography1} D¡¯Ariano, G.M., Lo Presti, P.: Quantum tomography for measuring experimentally the matrix elements of an arbitrary quantum Operation. Phys. Rev. Lett. \textbf{86}, 4195 (2001)
\bibitem{tomography2} D¡¯Ariano, G.M., Lo Presti, P., Paris, M.G.A.: Using entanglement improves the precision of quantum measurements. Phys. Rev.
Lett. \textbf{87}, 270404 (2001)
\bibitem{DLCZ}  Duan, L.M.,  Lukin, M.D., Cirac, J.T., Zoller, P.: Long-distance
quantum communication with atomic ensembles and linear optics. Nature \textbf{414}, 413-418 (2001)

\bibitem{telescope} Gottesman, D.,  Jennewein, T.,  Croke,S.: Longer-baseline telescopes using quantum repeaters.  Phys. Rev. Lett. \textbf{109}, 070503 (2012)
\bibitem{NLA1}  Ralph, T.C., Lund, A.P.: Nondeterministic noiseless linear
amplification of quantum systems. in Proceedings of the 9th
International Conference on Quantum Communication
Measurement and Computing, A. lvovsky, ed. (AIP, 2009),
pp. 155-160

\bibitem{Gisin} Gisin, N., Pironio, S., Sangouard, N.: Proposal for implementing
device-independent quantum key distribution based on a heralded qubit amplifier.
Phys. Rev. Lett. \textbf{105}, 070501 (2010)


\bibitem{Xiang} Xiang, G.Y., Ralph, T.C., Lund, A.P., Walk, N., Pryde, G.J.: Heralded
noiseless linear amplification and distillation of entanglement.
Nat. Photonics \textbf{4}, 316-319 (2010)

\bibitem{Curty} Curty, M., Moroder, T.: Heralded-qubit amplifiers for practical
device-independent quantum key distribution.  Phys. Rev. A \textbf{84}, 010304(R) (2011)

\bibitem{Pitkanen} Pitkanen, D., Ma, X., Wickert, R., van Loock, P., L\"{u}tkenhaus, N.:
Efficient heralding of photonic qubits with application to device-inpendent quantum key
distribution. Phys. Rev. A \textbf{84}, 022325 (2011)

\bibitem{heralded} Osorio, C.I., Bruno, N., Sangouard, N., Zbinden, H., Gisin, N., Thew,
R.T.: Heralded photon amplification for quantum communication.
Phys. Rev. A \textbf{86}, 023815 (2012)

\bibitem{heralded2}  Kocsis, S., Xiang, G.Y., Ralph, T.C., Pryde, G.J.: Heralded noiseless
amplification of a photon polarization qubit.  Nat. Phys. \textbf{9}, 23-28 (2012)

\bibitem{heralded3} Bruno, N., Pini, V., Martin, A., Verma, V.B., Nam, S.W., Mirin, R., Lita, A., Marsili, F., Krozh, B., Bussi\`{e}res, F., Sangouard, N., Zbinden, H., Gisin, N., Thew, R.: Heralded amplification of photonic qubits. Opt. Express \textbf{24}, 125 (2016)

\bibitem{Evan} Meyer-Scott, E., Bula, M., Bartkiewicz, K., $\check{C}$ernoch, A., Soubusta,
J., Jennewein, T., Lemr, K.: Entanglement-based liner-optical qubit amplifier.
Phys. Rev. A \textbf{88}, 012327 (2013)

\bibitem{ouyang} Ou-Yang, Y., Feng, Z.F., Zhou, L., Sheng, Y.B.: Linear-optical qubit amplification
with spontaneous parametric down-conversion source. Laser Phys. \textbf{26}, 015204 (2016)

\bibitem{Zhang} Zhang, S.L., Yang, S., Zou, X.B., Shi, B.S., Guo, G.C.: Protecting
single-photon entangled state from photon loss with noiseless linear amplification.
Phys. Rev. A \textbf{86}, 034302 (2012)

\bibitem{zhoul} Zhou, L., Sheng, Y.B.: Distilling single-photon entanglement from
photon loss and decoherence. J. Opt. Soc. Am. B \textbf{30}, 2737-2742 (2013)

\bibitem{sheng2} Sheng, Y.B., Ou-Yang, Y., Zhou, L., Wang, L.: Protecting single-photon
multi-mode W state from photon loss. Quant. Inf. Process. \textbf{13}, 1595-1605 (2014)

\bibitem{wangtj} Wang, T.J., Cao, C., Wang, C.: Linear-optical implementation of
hyperdistillation from photon loss. Phys. Rev. A \textbf{89}, 052303 (2014)

\bibitem{wangtj2}Wang, T.J., Wang, C.: High-efficient entanglement distillation from photon loss and decoherence. Opt. Express \textbf{23}, 31550 (2015)
\bibitem{ou} Ou-Yang, Y., Feng, Z.F., Zhou, L., Sheng, Y.B.: Protecting single-photon
entanglement with imperfect single-photon source.
Quant. Inf. Process. \textbf{14}, 635-651 (2015)

\bibitem{feng} Feng, Z.F., Ou-Yang, Y., Zhou, L., Sheng, Y.B.: Distillation of arbitrary single-photon entanglement assisted with polarized Bell states. Quant. Inf. Process. \textbf{14}, 3693-3710 (2015)

\bibitem{zhouamp} Zhou, L., Sheng, Y.B.: Recyclable amplification protocol for the single-photon entangled state. Laser Phys. Lett. \textbf{12}, 045203 (2015)

\bibitem{timebin}  Bruno, N., Pini,V.,  Martin, A.,  Korzh, B.,
 Bussi$\grave{e}$res, F.,  Zbinden, H.,  Gisin, N., Thew, R.: Heralded amplification of photonic qubits. Opt. Express \textbf{24}, 125-133 (2016)

\bibitem{Bennett1}  Bennett, C.H.  Bernstein, H.J.,  Popescu, S., Schumacher, B.: Concentrating partial entanglement by local operations. Phys.
Rev. A \textbf{53}, 2046-2052 (1996)

\bibitem{swapping1} Bose, S., Vedral, V., Knight, P.L.: Purification via entanglement swapping and conserved entanglement. Phys. Rev. A
\textbf{60}, 194 (1999)

\bibitem{swapping2} Shi, B.S., Jiang, Y.K., Guo, G.C.: Optimal entanglement purification via entanglement swapping. Phys.
Rev. A \textbf{62}, 054301 (2000)

\bibitem{zhao1} Zhao, Z., Pan, J.W., Zhan, M.S.: Practical scheme for entanglement concentration. Phys. Rev. A \textbf{64}, 014301
(2001)

\bibitem{Yamamoto1} Yamamoto, T., Koashi, M., Imoto, N.: Concentration and purification scheme for two partially entangled photon pairs. Phys. Rev. A \textbf{64}, 012304 (2001)
\bibitem{shengpra2} Sheng, Y.B., Deng, F.G., Zhou, H.Y.: Nonlocal entanglement concentration scheme for partially entangled multipartite systems with nonlinear optics. Phys. Rev. A \textbf{77}, 062325 (2008)
\bibitem{shengqic}  Sheng, Y.B., Deng, F.G., Zhou, H.Y.: Single-photon entanglement concentration for long distance quantum communication. Quant. Inf. \& Comput. \textbf{10}, 0272-0281 (2010)

\bibitem{shengpra3} Sheng, Y.B., Zhou, L., Zhao, S.M., Zheng, B.Y.: Efficient single-photon-assisted entanglement concentration for partially entangled photon pairs. Phys. Rev. A \textbf{85}, 012307 (2012)

\bibitem{shengwstateconcentration} Sheng, Y.B., Zhou, L., Zhao, S.M.: Efficient two-step entanglement concentration for arbitrary W states. Phys. Rev. A \textbf{85}, 044305 (2012)

\bibitem{bose} Paunkovi\'{c}, N., Omar, Y., Bose, S., Vedral, V.: Entanglement concentration using quantum statistics. Phys. Rev. Lett. \textbf{88}, 187903 (2002)

\bibitem{wangxb} Wang, X.B., Fan, H.: Entanglement concentration by ordinary linear optical devices without postselection. Phys. Rev. A \textbf{68},
060302 (2003)

 \bibitem{wangc2} Wang, C., Zhang, Y., Jin, G.S.: Entanglement purification and concentration of electron-spin entangled states using quantum-dot spins in optical microcavities. Phys. Rev. A \textbf{84},  032307 (2011)

\bibitem{dengpra} Deng, F.G.: Optimal nonlocal multipartite entanglement concentration based on projection measurements. Phys. Rev. A \textbf{85}, 022311 (2012)

\bibitem{yeliu} Xiong. W., Ye L.: Schemes for entanglement concentration
of two unknown partially entangled states with cross-Kerr nonlinearity. J. Opt. Soc. Am. B \textbf{28}, 2030-2037 (2011)


\bibitem{gubin} Gu, B.: Single-photon-assisted entanglement concentration of partially entangled multiphoton W states with linear optics. J. Opt. Soc. Am. B \textbf{29}, 1685-1689 (2012)

\bibitem{zhounoon} Zhou, L., Sheng, Y.B., Cheng, W.W., Gong, L.Y., Zhao, S.M.: Efficient entanglement concentration  for arbitrary less-entangled NOON states. Quant. Inf. Process. \textbf{12}, 1307-1320 (2013)

\bibitem{zhounoon2} Zhou, L., Sheng, Y.B.: Efficient entanglement concentration for arbitrary less-entangled
NOON state assisted by single photons, Chin. Phys. B \textbf{25}, 020308 (2016)

\bibitem{zhouqip2} Zhou, L.: Efficient entanglement concentration for electron-spin W state with the charge detection. Quant. Inf. Process. \textbf{12}, 2087-2101 (2013)
\bibitem{zhouoc} Zhou, L., Sheng, Y.B.: Efficient single-photon entanglement concentration
for quantum communications. Opt. Commun. \textbf{313},  217-222 (2014)


\bibitem{photonefficiency}  D'Auria, V., Lee, N., Amri, T.,  Fabre, C., Laurat, J.: Quantum decoherence of single-photon counters. Phys. Rev. Lett. \textbf{107}, 050504 (2011)
\bibitem{photonefficiency1}   Lita, A.E., Miller, A.J., Nam, S.W.: Counting near-infrared single-photons with 95\% efficiency. Opt. Express \textbf{16}, 3032-3040 (2008)

\end{thebibliography}
\end{document}